# Incorporation of Eye-Tracking and Gaze Feedback to Characterize and Improve Radiologist Search Patterns of Chest X-rays: A Randomized Controlled Clinical Trial

*Carolina Ramirez-Tamayo[1], Syed Hasib Akhter Faruqui[2], Stanford Martinez[1], Angel Brisco[6], Nicholas Czarnek[4], Adel Alaeddini[1], Jeffrey R. Mock[5], Edward J. Golob[5], Kal L. Clark[3]*

[1] Department of Mechanical Engineering, The University of Texas at San Antonio, Texas, United States.

[2] Department of Radiology, Northwestern University, Chicago, Illinois.

[3] Department of Radiology, University of Texas Health Science Center at San Antonio, Texas, United States.

[4] Independent Researcher, Anaheim, California, United States.

[5] Department of Psychology, The University of Texas at San Antonio, Texas, United States.

[6] Independent Researcher, Corrientes, Argentina.

**Abstract:**

Diagnostic errors in radiology often occur due to incomplete visual assessments by radiologists, despite their knowledge of predicting disease classes. This insufficiency is possibly linked to the absence of required training in search patterns. Additionally, radiologists lack consistent feedback on their visual search patterns, relying on ad-hoc strategies and peer input to minimize errors and enhance efficiency, leading to suboptimal patterns and potential false negatives.

This study aimed to use eye-tracking technology to analyze radiologist search patterns, quantify performance using established metrics, and assess the impact of an automated feedback-driven educational framework on detection accuracy. Ten residents participated in a controlled trial focused on detecting suspicious pulmonary nodules. They were divided into an intervention group (received automated feedback) and a control group.

Results showed that the intervention group exhibited a 38.89% absolute improvement in detecting suspicious-for-cancer nodules, surpassing the control group's improvement (5.56%, p-value=0.006).

Improvement was more rapid over the four training sessions (p-value=0.0001). However, other metrics such as speed, search pattern heterogeneity, distractions, and coverage did not show significant changes.

In conclusion, implementing an automated feedback-driven educational framework improved radiologist accuracy in detecting suspicious nodules. The study underscores the potential of such systems in enhancing diagnostic performance and reducing errors. Further research and broader implementation are needed to consolidate these promising results and develop effective training strategies for radiologists, ultimately benefiting patient outcomes.

1. **Introduction:**

Lung cancer is responsible for 1.59 million deaths per year globally [1]. Early-stage diagnosis relies on radiological chest images, which radiologists visually inspect to identify key findings and make recommendations on further medical management. The collection of visual inspection strategies used by a radiologist is colloquially termed "search pattern." During the visual inspection process, radiologists are unfortunately prone to perception error with false negative estimates approaching 33% for abnormal (e.g., mass, effusion, nodules, etc.) chest exams, delaying or preventing patients from entering lung cancer prevention and treatment programs [2]. Kundel et al. [3] classified perceptual errors into three classes: *search errors*, for which the radiologist's eye gaze does not cover the abnormality; *recognition errors,* for which the radiologist's eye gaze covers the abnormality, but the radiologist remains unaware of its presence; and *decision errors*, for which the radiologist fixates on the abnormality, but incorrectly interprets the finding as no abnormality or the wrong type of abnormality. Renfrew et al. [4] investigated 182 radiology cases, revealing that around 69% of errors are related to search and recognition errors. Currently, available error mitigation strategies do not directly address search and recognition errors by the original radiologist but instead rely on double reporting by a second radiologist, continuing medical education, and artificial intelligence deployment [5]. Although most radiology literature supports the notion that errors are primarily search and recognition, Manning et al. [6] found that most errors are due to decision errors rather than search and recognition.

Busby et al. found that additional sources of false-negative error are related to temporary human environmental variables and include workplace interruption and fatigue and further suggested that reducing workplace interruptions prevents disruption of the cognitive memory of the original task and improves detection performance [7]. As an example, Figure 1 illustrates the search pattern of a radiologist who experienced two interruptions during an X-ray assessment. When the radiologist had an interruption, there was a gap between the start and stop point which could increase the chance of mistakes. To preserve the continuity of the search pattern, an optimal scenario would involve the radiologists being able to seamlessly resume their reading from the point where they were interrupted.

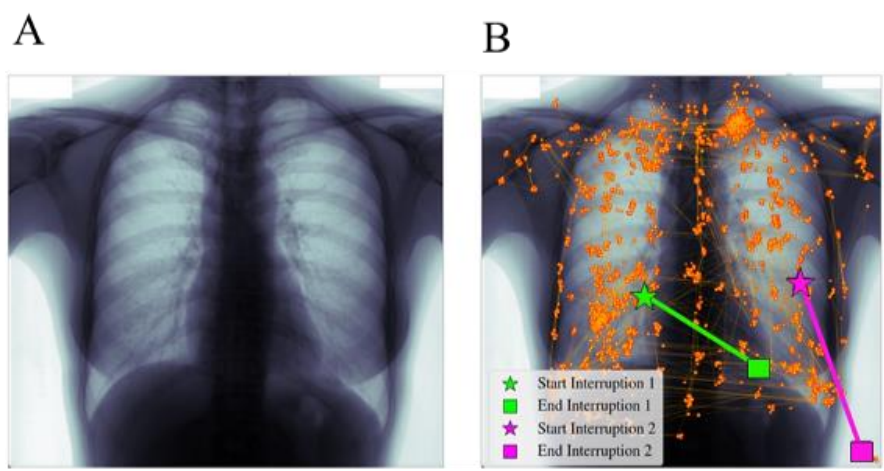

*Figure 1. (a) X-ray without superimposed eye movement search data; (b) Interruptions made by a radiologist. the orange dots correspond to the positions of the eyes during the X-ray scanning process. The green and purple lines indicate the occurrence of interruptions, with the star denoting the initiation of the interruption, and the square marking the point at which the radiologist resumes reading the X-ray. The analysis demonstrates that interruptions have a discernible impact on the continuity of the search pattern employed by the radiologist during the examination.*

Bruno et al. [8] proposed the implementation of Lean Six Sigma techniques, a methodology combining Lean principles for waste reduction and Six Sigma methodologies for process variation reduction, in radiological assessments. This approach aims to optimize efficiency and accuracy in radiology practices, particularly through the integration of assistive tools. Moreover, the authors advocate the incorporation of feedback modules into radiologists' training programs to effectively address and diminish diagnostic errors in their practice. The combination of Lean Six Sigma techniques and feedback modules represents a holistic approach to enhancing radiological proficiency and streamlining diagnostic processes.

In the realm of medical education, a systematic review of eye-tracking technology [8] found that several studies have verified the utility of eye-tracking methodology in detecting visual gaze pattern discrepancies among clinicians in radiology. Kundel et al. [3] had previously recognized the potential of eye-tracking methodology in identifying inter-observer differences in radiology during the early 1990s. More recently, Rubin et al. [9] conducted a study to track the time-varying gaze paths of radiologists, while Wood et al. [10] employed a cohort study to uncover variations in visual search strategies between experienced professionals and novices while examining X-rays. Tourassi et al. [11] also probed the relationship between image content, human perception, cognition, and diagnostic errors in mammogram interpretations. However, to the best of our knowledge, no educational feedback and intervention framework utilizing eye-tracking technology specifically tailored for the chest radiology domain has been established.

In line with the ongoing emphasis on continuous improvement, our hypothesis posits that offering regular feedback to radiologists concerning their performance, as measured by eye-tracking technology, could lead to enhancements in search pattern quality, encompassing both efficiency and accuracy. Should this approach yield positive results, it could be effectively integrated into radiology training, thereby mitigating error prevalence and serving as a valuable metric to gauge a radiologist's preparedness for further education, licensing, or credentialing.

Figure 2 shows the proposed schema behind this work. Eye-tracking data is first collected from radiologists, including experts and novices. Next, we analyze the collected data and compute proposed metrics. The intervention group of the study is provided with the proposed metrics as feedback. At the end of the 4 sessions, we assess overall changes in performance compared to the subject's baseline.

The goal of this paper is to provide a system to characterize radiologist search patterns using eye-tracking, test performance using commonly accepted metrics, and evaluate the potential of a unique automated feedback-driven educational framework to enhance radiologist detection performance.

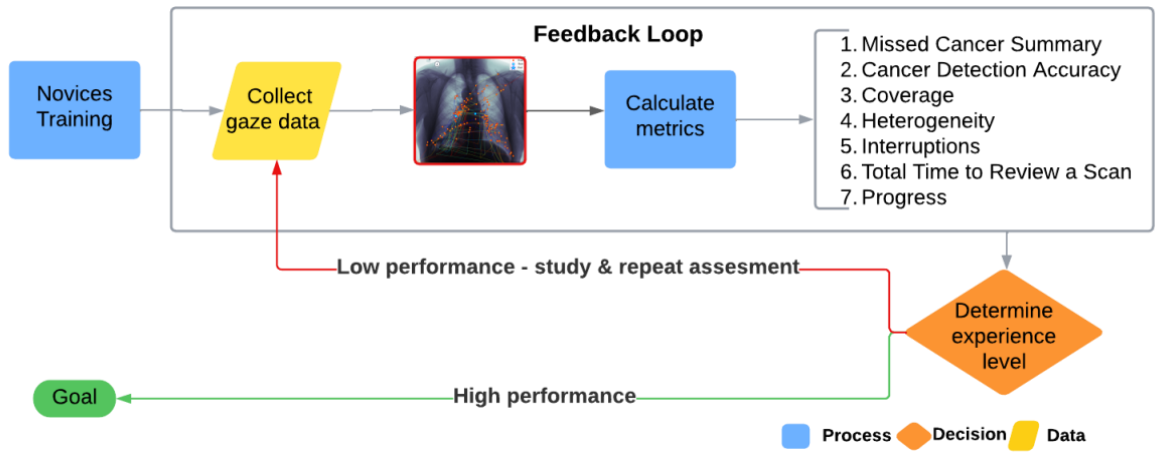

*Figure 2. The eye-tracking data from a radiologist were collected and processed, then analyzed to generate key performance metrics, which were provided to the radiologists in the intervention group at the end of each session. After each session, we assessed if performance improved.*

2. **Materials and Methods:**

2.1. **Experiment Description:**

The study design was prospective, controlled, longitudinal, block-randomized, and IRB approved. Inclusion criteria included active radiology residents and faculty in the author's [KLC] academic radiology department. The study was powered to detect a 10% mean absolute difference between the control and the intervention group (alpha = 0.05, power = 80%, enrollment ratio = 1). Twenty-one subjects were enrolled: 6 faculty, one of whom had completed a chest fellowship, and 15 residents. Subjects were randomly assigned to the intervention group with a probability of 0.5, while the remainder were assigned to the control group. Separate enrollment blocks for faculty and residents were implemented to ensure expertise balance. We used sequentially numbered containers to implement the random allocation sequence. The generation of random allocation sequence, participant enrollment, and assignment of participants to interventions was managed by the principal investigator [KLC]. Each study subject completed four one-hour sessions (session 1: baseline, session 2-4: follow-up) in a radiology reading room over a 1-year study period. During the one-hour session, the following activities were conducted: calibration, labeling instruction, review and dictation of case 1, labeling of case 1, review and dictation of case 2, and so forth until the session's conclusion. The primary outcome measure was the accuracy of positive sample detection (sensitivity), and the secondary measures were exam coverage, search pattern heterogeneity, number of search pattern interruptions, and time to review a normal scan. Fifteen subjects completed the study: 3 faculty and 12 residents. Two residents did not complete their first session and were excluded. Since only 3 faculty completed the study these were excluded from the analysis. Therefore, 5 resident subjects each in the intervention and control groups were used for analysis.

Control group subjects received no feedback between sessions, while the intervention group received feedback after Sessions 1-3, with the feedback restricted solely to the information contained within the report. This design aimed to establish a fully autonomous feedback system for the participants. The inter-session interval spanned 4 days, ensuring that feedback was provided with a minimum lead time of one day before the subsequent session. Feedback was delivered via email and provided in print form immediately before Sessions 2-4. The performance report provided participants with feedback categorized into six metrics, elucidating the quality of their readings.

During each assessment Session, subjects evaluated 33 cases: 18 cases with a single pulmonary nodule (suspicious for cancer), 9 normal cases, and 6 distractor cases (pneumothorax, cardiomegaly, and consolidation). All subjects were blinded to the study objective of pulmonary nodule detection during enrollment to mitigate detection bias. Distractor cases were included to further mitigate detection bias, preventing subjects from deducing the purpose of the experiment due to the proportion of pulmonary nodule cases. Nodule and normal cases were derived from the Shiraishi 2000 JSRT (Japanese Society of Radiological Technology) chest radiograph dataset [12], which includes 154 abnormal chest radiographs with 5 degrees of subtlety from level 1 (extremely subtle) to level 5 (obvious) and pathology follow-up, and 93 normal chest radiographs. Distractor cases were derived from the VinDr chest radiograph dataset [13]. Three sets of six nodule cases from the JSRT dataset [14], one set each from the intermediate difficulty levels (2, 3, & 4), and one set of nine normal cases from the JSRT dataset were randomly sampled without replacement. The primary focus of this investigation revolves around evaluating the proficiency of radiologists in identifying "nodules of intermediate difficulty." Two cases each of pneumothorax, cardiomegaly, and consolidation from the VinDr dataset were randomly sampled without replacement. No scan was reviewed twice by a study subject during a session. The cases were not repeated block to block; all study subjects reviewed the same four sets of cases, one for each of the 4 sessions, with exams presented in the same order for each study participant.

The subjects in this study were asked to voice dictate their radiograph findings into a microphone as they would normally do in a standard-of-care assessment. Next, each radiologist outlined any abnormal areas mentioned during a subsequent dictation phase which was initiated via a key press on the keyboard. Subjects were given the liberty to designate multiple locations within an image while remaining unaware that only one nodule existed per positive case. Subjects completed a survey before and after each session to assess fatigue and predictions on self and peer performance. A custom software tool was used to conduct the study which allowed automatic display of the study images and capture of timestamped bilateral gaze, bilateral pupil, head pose, voice, annotation, and image display configuration data. Gaze, pupil, and head pose data was captured using the Tobii 5L system at 30 Hz [15].

## 2.2. Metrics

In our implementation, we incorporate a feature set, partially derived from Van der Gijp et al. [16] and Liu et al. [17] to develop a comprehensive quantitative radiologist search pattern performance assessment. To quantify radiologists' consistency in scanning normal chest X-rays (CXRs), we developed an additional feature metric: search pattern heterogeneity. Our proposed performance assessments represent a technique to illustrate the relationship between visual perception and our primary and secondary outcomes. Table 1 lists the suggested performance metrics, their descriptions, and our corresponding hypotheses about how performance may change in the following four sessions.

*Table 1. Proposed Performance Assessment Metrics.*

| Metric | Description | Expected change after four sessions with feedback |
| --- | --- | --- |
| Cancer detection accuracy | Numerical abnormality detection sensitivity | Increase |

| Coverage | Percentage of the lungs visualized during the assessment | Increase |
|---|---|---|
| Heterogeneity | Search pattern dissimilarity among normal scans | Decrease |
| Response to interruptions | Number of interruptions occurring during scan review | Decrease |
| Total time to review a normal scan | How long does a radiologist take to review a scan compared to peers | Decrease |

In this study, the participants underwent a comprehensive evaluation based on a set of diverse metrics. The outcomes of this evaluation were presented to each participant in a printed report. Notably, the cancer summary assessment was depicted using a heatmap, which integrated their eye-tracking data. The heatmap allowed for a succinct representation of both accurate and erroneous identifications of cancerous features in each X-ray image, along with highlighting regions of heightened visual focus. Additionally, the participant's performance on each of the other metrics was gauged through the assignment of scores. To facilitate comparative analysis, graphical representations of these scores were provided, enabling participants to assess their performance relative to both expert diagnosticians, stratified by their experience, and peers of similar proficiency.

2.2.1. **Cancer Detection Summary:**

For each radiologist session, heatmaps were created using averaged gaze data collected across all evaluations to provide summative feedback to radiologists about the areas which they regularly inspected and those which they often failed to evaluate and/or detect (see Figure 3 (a)). The driving hypothesis for the generation of these visuals is that most abnormalities are missed not because of decision error, but rather due to search error. Summary heatmaps were created by spatially mapping gaze data collected during each scan onto a 2D grid, smoothing it using a radial filter, binarizing it, overlaying, and summing all binarized scans to create the final heatmap. The brighter areas on the heatmap correspond to areas that radiologists more often evaluated. This process was done to provide feedback on areas that radiologists regularly inspected and missed during their evaluations. Figure 3 (a) shows the provided heatmap and (b) the location of the nodule, and (c) eye gaze with the abnormality location.

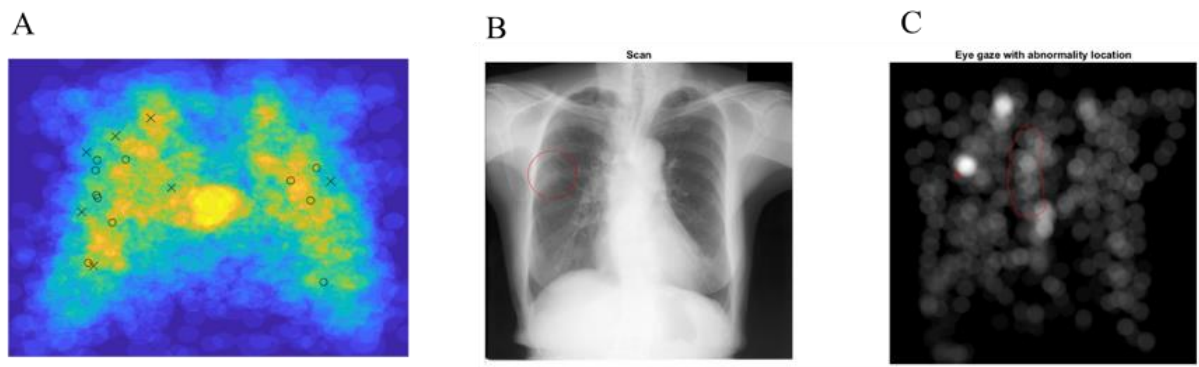

*Figure 3. (A) Overall performance of a radiologist subject with 61% sensitivity based on gaze data. The figure summarizes the subjects' misses and hits and the coverage and intensity of gaze visitation. A false negative is marked as (X) while a true positive is marked as (O); (B) True positive or false negative: The figure presents the location of the abnormality as a red circle; and (C) An illustration of the radiologist's gaze while reading the X-ray, and whether they detected the abnormality. In this case, the radiologist marked the patient's right suprahilar region as abnormal but failed to detect the nodule in the patient's right midlung. However, it is interesting to note that the radiologist did visually interrogate the nodular area, as shown by the high intensity of the heatmap near the abnormality marker.*

2.2.2. **Cancer Detection Accuracy**

The proportion of correct assessments was averaged throughout each session for each radiologist. Each radiologist was provided with accuracy for the session, along with summaries of faculty and resident accuracy to demonstrate relative performance compared to peers. Figure 4 (a) shows the accuracy of a Resident in session 1 (as provided in the performance report).

2.2.3. **Coverage**

We defined coverage as the proportion of the lung reviewed by each radiologist. Coverage was calculated using a denominator equal to the number of pixels comprising the lungs (cumulatively across all scans) and a numerator equal to the number of pixels within the lungs reviewed by the radiologist as measured by the eye-tracker (also cumulative). Figure 4 (b) shows the probability distribution of percentage coverage of CXR's area of interest of a Resident in Session 1 (as provided in the performance report).

2.2.4. **Heterogeneity**

Van Der Gijp et al. and Kok et al. [16], [18] have researched eye-tracking in radiography to understand how visual search patterns are linked to high perceptual performance and how distinct types of images influence how they are perceived by radiologists. The researchers have also studied how comparable radiologists' search patterns are among normal instances and how they compare to colleagues at the expert level. This is referred to as the "heterogeneity problem" and is measured by the consistency of their search pattern and how it affects their accuracy.

We employed Dynamic-Time-Warping (DTW) [19], which measures the similarity of a set of time series, all of which may have a variable duration, to calculate heterogeneity. The time series analyzed for this research were the x and y positions recorded by the eye-tracker over time for each image [19]. The resulting heterogeneity matrix shows the similarity (low heterogeneity value) or dissimilarity (high heterogeneity value) for each pair of images reviewed. We calculated the average heterogeneity along with the standard deviation for each radiologist and provided a comparison to the faculty author's (KLC) heterogeneity mean and standard deviation. Figure 4 (c) depicts the resulting matrix from comparing all search patterns from each session's nine normal exams, with 0 and 8 representing the first and last normal images Figure 4 (d) depicts a PDF comparing the subject's heterogeneity to that of an expert.

2.2.5. **Interruptions**

Busby et al. [7] presented workplace interruptions as an obstacle to minimizing cognitive and perceptual errors. Subjects' performance in terms of accuracy and sensitivity may improve across sessions with fewer interruptions. Radiologists who are interrupted during X-ray evaluations may lose the continuity of the search pattern (shown in image 1(a)), increasing the potential for a cognitive error. Based on [20], we considered any distractions that averted the radiologist's gaze away from a scan for more than 500 milliseconds to be an interruption and aversions less than 500 milliseconds to blink. Uniform reading conditions were maintained across all participants, and any interruptions encountered during the reading process were purely intrinsic in nature, without artificial generation. Figure 4 (e) depicts the number of interruptions experienced by a resident in session 1 like that shown in Figure 1(a).

2.2.6. **Total Time to Review a Scan.**

Van Der Gijp et al [16] summarized literature regarding the average time it takes professionals and beginners to scan an X-ray and concluded that there was a large variance in average expert viewing times across trials, ranging from 4 to nearly 45 seconds. Based on the author's experience, novices regularly take 1.5-2.5 times as long to review scans. We hypothesize that as radiologists acquire experience, it will take them less time to review a scan properly, making them more efficient. We provide radiologists with their total time to review a scan. Figure 4 (f) depicts the overall time it took a resident to review a scan during their first session across different X-ray images.

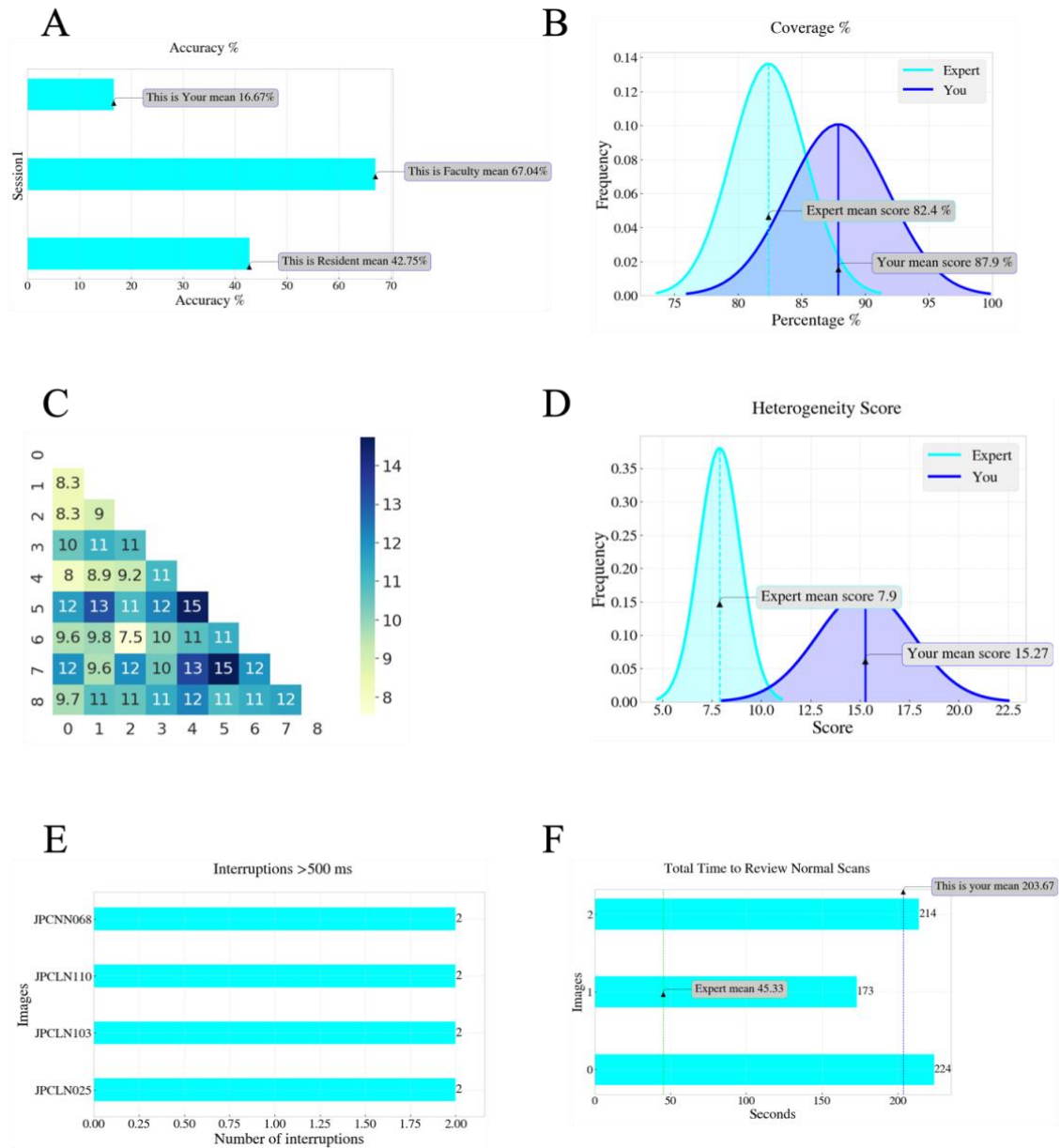

*Figure 4. Feedback provided to radiologists in the performance report: (A) Accuracy of a resident in the first session compared to their peers and faculty members (B) Probability distribution of percentage coverage of CXR's area of interest compared to an expert; (C) Dynamic Time Warping matrix comparing homogeneity of search patterns in session 1 images for a resident. (Smaller scores show more homogeneity); (D) Probability distribution of heterogeneity score compared to an expert (The lower the mean value the better); (E) Interruptions made by a resident in session; (F) Time to scan the CXR per session compared to an expert.*

## 3. Discussion

Figure 5 illustrates the average change in performance metrics across the four sessions for both the intervention and control groups. The results indicate a significant enhancement in nodule detection accuracy (sensitivity) for the intervention group, with an average absolute improvement of 38.89% over the sessions (112.90% relative improvement), in contrast to the control group's 5.56% absolute improvement. The mixed-effect ANOVA test summarized in

Table 2 supports this finding, revealing a main effect of group ($p = 0.006$) and intervention session ($p = 0.0001$). Conversely, the other performance metrics, including coverage, heterogeneity, interruption, and time, did not demonstrate statistically significant changes based on the mixed-effect ANOVA test, as indicated in

Table 2. Nonetheless, these metrics displayed slight absolute improvement compared to the control group (See Also Figure 5). The non-significant ANOVA results for these metrics may be attributed to the relatively high within-group variation, particularly for the intervention group, as they acclimate to the new feedback framework. Appendix B presents individual analyses for two randomly selected subjects, highlighting the variance of responses within the intervention group and offering insights into the report generation and analysis process.

While the overall results support the hypothesis that feedback provision improves radiologists' performance, the study is not without its limitations. The framework's reliance on eye-tracking data introduces sensitivity to noise and potential bias in cases of data errors. Efforts were made to simulate standard radiologist environments closely for maximum reproducibility, but the lack of enforcing a prescribed distance from the screen may have affected gaze calibration temporarily. Additionally, raw gaze data, as opposed to fixations, were used for all analyses, which may have resulted in an overestimation of the coverage metric compared to fixation-based analyses.

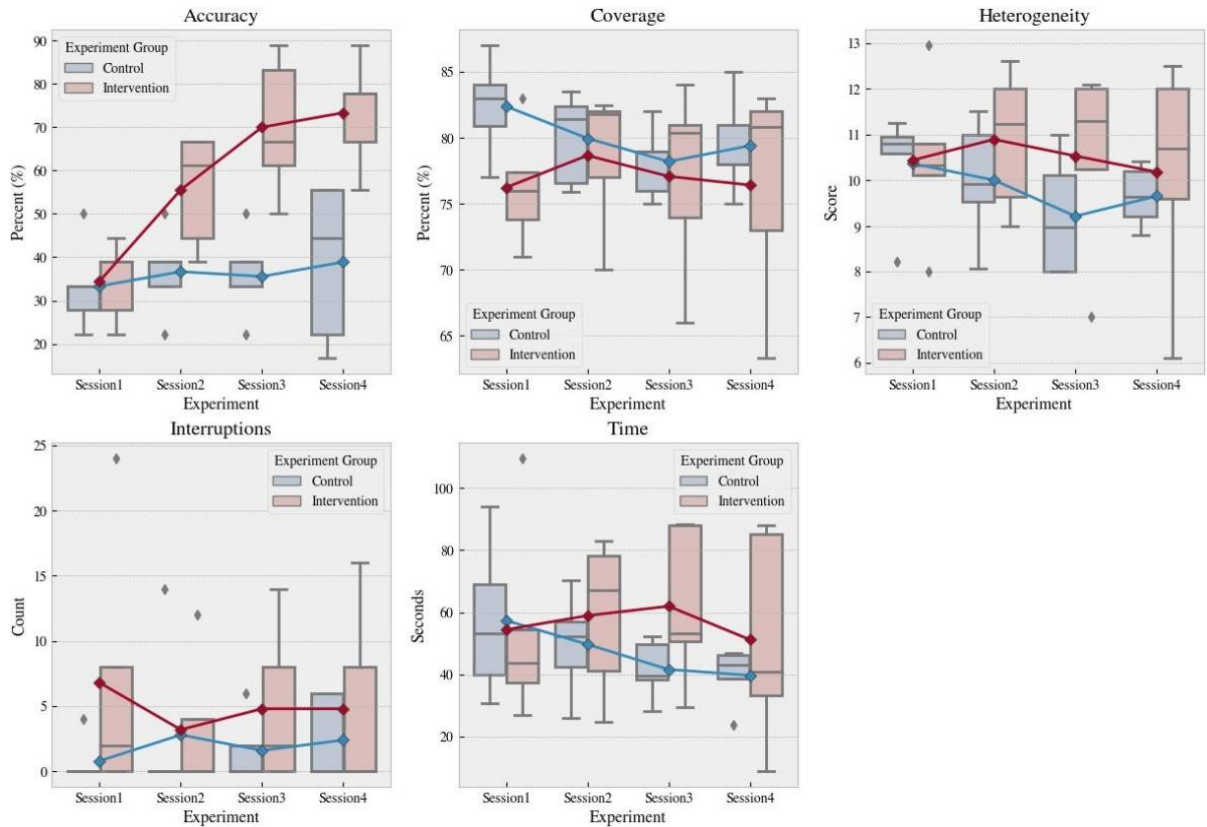

*Figure 5. The progression of performance metrics (accuracy, coverage, heterogeneity, interruption, and time) in control vs intervention groups across the four sessions. The accuracy metric shows a meaningful improvement over the 4 sessions. Other metrics including coverage, homogeneity, interruptions, and time do not show a meaningful change.*

*Table 2. Summary results for the study group based on mixed-effect ANOVA*

| Variable | Expected Behavior | Baseline Performance | | Absolute Improvement | | p-value | | |
|---|---|---|---|---|---|---|---|---|
| | | Intervention Group | Control Group | Intervention Group | Control Group | Group | Session | Interaction |
| Accuracy | Increase | 34.44% | 33.33% | 38.89% | 5.56% | 0.006 | 0.0001 | 0.001 |
| Coverage | Increase | 76.24% | 82.38% | 0.18% | -2.98% | 0.347 | 0.395 | 0.204 |
| Heterogeneity | Decrease | 10.43 | 10.36 | -0.26 | -0.71 | 0.388 | 0.639 | 0.738 |
| Interruptions | Decrease | 6.80 | 0.80 | -2.00 | 1.60 | 0.386 | 0.944 | 0.312 |
| Time | Decrease | 54.41 | 57.36 | -3.21 | -17.67 | 0.454 | 0.524 | 0.486 |

## 4. Conclusion

In this study, we present an educational framework designed to enhance radiologists' skills and performance. The proposed Lean Six Sigma-based methodology, integrated with feedback modules, aimed to optimize efficiency and accuracy in radiological assessments. The intervention group demonstrated a significant improvement in nodule detection accuracy, indicating the potential efficacy of the framework. While other performance metrics did not exhibit statistically significant changes, their marginal improvements warrant further investigation.

We acknowledge certain limitations in our study, such as the sensitivity of the eye-tracking data and the potential for variability due to user movements. As future steps, we plan to expand the data collection during routine clinical care to increase sample size and mitigate potential Hawthorne effects. Additionally, we intend to assess specificity and inter-subject variability, involving a larger pool of radiologists, including faculty members. The evaluation of data presentations and responses to different feedback formats will also be explored to optimize performance outcomes.

Our educational framework offers a valuable tool for radiologists, providing accurate self-assessment and skill-based breakdowns without additional time investment. Unlike existing subjective methods, this automated approach fosters a constructive learning environment, empowering radiologists to mitigate errors and enhance their diagnostic abilities effectively. Overall, this framework holds promise in augmenting radiological practices and reducing diagnostic errors in clinical settings.

**Take Home Points:**

1) Lung cancer results in 1.59 million deaths per year globally. Screening chest radiography has a false negative percentage of 33% for abnormality detection, which prevents patients from receiving early and accurate treatment. Perception-related errors made by radiologists comprise most of them.

2) There is not a gold-standard search pattern radiologists should follow, and the healthcare system is limited in tools to educate radiologists.

3) Using eye-tracking data, we develop an educational framework that evaluates radiologists' perceptual performance and compares it to peers, establishing a competitive environment in which errors are seen as opportunities to improve. Application of this framework in the intervention group resulted in a 38.89% absolute improvement in suspicious nodule detection (group effect p-value = 0.006).

**Summary Sentence**: We propose an educational framework to assist radiologists in improving their skills and performance based on collected eye-tracking data. This educational framework may help radiologists increase accuracy by reducing perceptual errors.

**Author Contribution**:

K.C., A.A., J.M., and E.G., developed and proposed the concept. Data collection (collection process and relevant section) and medical knowledge was provided by K.C. Psychological analysis and perception-related error research were done by J.M. and E.G. Performance metrics were coded by C.R. and N.C. Data Analysis and methodology design were conducted by C.R, S.H.A.F, N.C., and K.C., and the manuscript was written by C.R., S.H.A.F.; All authors reviewed the manuscript. All authors read and approved the final manuscript.

**Disclosure**:

Conflict of Interest Disclaimer: Kal Clark & Nicholas Czarnek are owners of a company that incorporates eye-tracking for machine learning applications. All other authors of this paper declare that they have neither conflicts of interest nor financial or personal relationships with any individuals or organizations that could inappropriately influence their work.

**Ethics approval and consent to participate**:

The Ethics approval and consent to participate is provided in IRB 19-533H. The study was performed under a protocol approved by the University of Texas, San Antonio Institutional Review Board, and participants gave written informed consent before testing. The trial protocol is available upon request to the IRB.

**Data Availability**:

Images and annotations were obtained from the NIH Clinical Center and can be downloaded at https://nihcc.app.box.com/v/ChestXray-NIHCC. The IRB restricts sharing of the gaze data. In case further information, please contact Kal L. Clark (clarkkl@uthscsa.edu).

**Funding**:

This research is funded by San Antonio Medical Foundation (SAMF), PI: Kal Clark (UTHSA), Edward Golob (UTSA).

## 5. References

[1] P. A. Groome *et al.*, "The IASLC Lung Cancer Staging Project: validation of the proposals for revision of the T, N, and M descriptors and consequent stage groupings in the forthcoming (seventh) edition of the TNM classification of malignant tumours," *J. Thorac. Oncol. Off. Publ. Int. Assoc. Study Lung Cancer*, vol. 2, no. 8, Art. no. 8, Aug. 2007, doi: 10.1097/JTO.0b013e31812d05d5.

[2] S. Waite *et al.*, "Analysis of Perceptual Expertise in Radiology – Current Knowledge and a New Perspective," *Front. Hum. Neurosci.*, vol. 13, p. 213, Jun. 2019, doi: 10.3389/fnhum.2019.00213.

[3] H. L. Kundel, C. F. Nodine, and D. Carmody, "Visual Scanning, Pattern Recognition and Decision-making in Pulmonary Nodule Detection," *Invest. Radiol.*, vol. 13, no. 3, Art. no. 3, Jun. 1978.

[4] D. L. Renfrew, E. A. Franken, K. S. Berbaum, F. H. Weigelt, and M. M. Abu-Yousef, "Error in radiology: classification and lessons in 182 cases presented at a problem case conference.," *Radiology*, vol. 183, no. 1, Art. no. 1, Apr. 1992, doi: 10.1148/radiology.183.1.1549661.

[5] R. Murphy, A. Slater, R. Uberoi, H. Bungay, and C. Ferrett, "Reduction of perception error by double reporting of minimal preparation CT colon," *Br. J. Radiol.*, vol. 83, no. 988, Art. no. 988, Apr. 2010, doi: 10.1259/bjr/65634575.

[6] D. J. Manning, S. C. Ethell, and T. Donovan, "Detection or decision errors? Missed lung cancer from the posteroanterior chest radiograph," *Br. J. Radiol.*, vol. 77, no. 915, Art. no. 915, Mar. 2004, doi: 10.1259/bjr/28883951.

[7] L. P. Busby, J. L. Courtier, and C. M. Glastonbury, "Bias in Radiology: The How and Why of Misses and Misinterpretations," *RadioGraphics*, vol. 38, no. 1, Art. no. 1, Jan. 2018, doi: 10.1148/rg.2018170107.

[8] M. Bruno, E. Walker, and H. Abujudeh, "Understanding and Confronting Our Mistakes: The Epidemiology of Error in Radiology and Strategies for Error Reduction," *RadioGraphics*, vol. 35, pp. 1668–1676, Oct. 2015, doi: 10.1148/rg.2015150023.

[9] G. D. Rubin *et al.*, "Pulmonary nodules on multi-detector row CT scans: performance comparison of radiologists and computer-aided detection," *Radiology*, vol. 234, no. 1, pp. 274–283, Jan. 2005, doi: 10.1148/radiol.2341040589.

[10] G. Wood, K. M. Knapp, B. Rock, C. Cousens, C. Roobottom, and M. R. Wilson, "Visual expertise in detecting and diagnosing skeletal fractures," *Skeletal Radiol.*, vol. 42, no. 2, pp. 165–172, Feb. 2013, doi: 10.1007/s00256-012-1503-5.

[11] G. Tourassi, S. Voisin, V. Paquit, and E. Krupinski, "Investigating the link between radiologists' gaze, diagnostic decision, and image content," *J. Am. Med. Inform. Assoc. JAMIA*, vol. 20, no. 6, pp. 1067–1075, 2013, doi: 10.1136/amiajnl-2012-001503.

[12] J. Shiraishi *et al.*, "Development of a digital image database for chest radiographs with and without a lung nodule: receiver operating characteristic analysis of radiologists' detection of pulmonary nodules," *AJR Am. J. Roentgenol.*, vol. 174, no. 1, Art. no. 1, Jan. 2000, doi: 10.2214/ajr.174.1.1740071.

[13] H. Q. Nguyen *et al.*, "VinDr-CXR: An open dataset of chest X-rays with radiologist's annotations," *Sci. Data*, vol. 9, no. 1, Art. no. 1, Jul. 2022, doi: 10.1038/s41597-022-01498-w.

[14] "JSRT Database | Japanese Society of Radiological Technology," Jan. 18, 2023. http://db.jsrt.or.jp/eng.php (accessed Jan. 17, 2023).

[15] "Tobii Eye Tracker 5L | Engineered for innovation - Tobii," Nov. 02, 2022. https://www.tobii.com/products/integration/pc-and-screen-based/tobii-eye-tracker-5l (accessed Nov. 02, 2022).

[16] A. van der Gijp *et al.*, "How visual search relates to visual diagnostic performance: a narrative systematic review of eye-tracking research in radiology," *Adv. Health Sci. Educ. Theory Pract.*, vol. 22, no. 3, Art. no. 3, Aug. 2017, doi: 10.1007/s10459-016-9698-1.

[17] P. T. Liu, C. D. Johnson, R. Miranda, M. D. Patel, and C. J. Phillips, "A Reference Standard-Based Quality Assurance Program for Radiology," *J. Am. Coll. Radiol.*, vol. 7, no. 1, pp. 61–66, Jan. 2010, doi: 10.1016/j.jacr.2009.08.016.

[18] E. M. Kok, A. B. H. De Bruin, S. G. F. Robben, and J. J. G. Van Merriënboer, "Looking in the same manner but seeing it differently: Bottom-up and expertise effects in radiology," *Appl. Cogn. Psychol.*, vol. 26, pp. 854–862, 2012, doi: 10.1002/acp.2886.
[19] "DTW for Python - The DTW suite," Nov. 02, 2022. https://dynamictimewarping.github.io/python/ (accessed Nov. 02, 2022).
[20] A. Królak and P. Strumiłło, "Eye-blink detection system for human–computer interaction," *Univers. Access Inf. Soc.*, vol. 11, no. 4, Art. no. 4, Nov. 2012, doi: 10.1007/s10209-011-0256-6.

**Appendices:**

**Appendix A:**

To quantify the performance of a radiologist using eye-tracking, we identified several established metrics for predicting expert radiologist performance. In 2017, Van der Gijp et al. [16] developed a consensus set of features in radiology, presented in Table 1, which consists of nine quantitative features that can be measured experimentally and correlate with the level of experience.

*Table A.1: List of Consensus Features to Quantify a Radiologist's Performance. See [18] for Descriptions.*

| Attribute (per trial) | Association with a high level of expertise (number of studies with significant results) |
|---|---|
| Total Time | Decrease (10) |
| Time to the first fixation | Increase (1) or decrease (1) |
| Fixation duration on an area of interest (AOI) | Decrease (4) |
| Dwell time ratio | Increase (2) or decrease (1) |
| Total number of fixations | Increase (2) or decrease (2) |
| Number of fixations on AOI (Area of Interest) | Increase (1) or decrease (1) |
| Saccade length | Increase (2) or decrease (1) |
| Image coverage | Decrease (2) or increase (2) |

Similarly, Liu et al [17], created a comprehensive radiology performance assurance (QA) system that assesses radiological operations and interpretations compared to accuracy rates (the proportion of correct interpretations to all interpretations) as performance indicators, which measure performance and can be considered a state-of-the-art feature.

**Appendix B: Individual Observations**

Table B.1 shows the summary of the changes of two subjects from the intervention group (at the end of each session; compared to baseline). At the end of the designated sessions (four sessions), we compared both their performances (baseline vs. the last session). Compared to their baselines, both subjects showed improvement in almost all the metrics.

*Table B.1 – Changes in performance metrics after receiving continuous feedback*

|  | Percentage Change in Performance Compared to Baseline | | | |
| --- | --- | --- | --- | --- |
| **Metrics** | **Session 2** | **Session 3** | **Session 4** | **Results** |
| Sensitivity | 50.00% | 37.50% | 75.00% | Improved |
| Coverage | 5.68% | 3.88% | 4.39% | Improved |
| Heterogeneity | 16.67% | 4.63% | -0.93% | Improved |
| Interruptions | 0.00% | 0.00% | 0.00% | No change |
| Total Time | 23.53% | -2.02% | -25.00% | Improved |

Subject 12, a resident in their first session, misdiagnosed 10 out of 18 CXRs. While on average, they took around 54.4 seconds to scan the CXRs. However, in the next few sessions, the subject showed steady improvement. The misdiagnosis rate for the subject for the final session was 4 out of 18, which is a 75.00% improvement from the baseline. We didn't notice a significant improvement in subjects' heterogeneity and coverage scores over the sessions (Figures B.1 & B.2). Although the changes are on the side of improvement, we think more sessions are needed on this subject to further improve their performance. One interesting observation we noticed in the second session (the first session after they received feedback) was that the subject took longer to scan the CXR than in other sessions. As a result, the heterogeneity score of the subject for this session was higher than the other sessions. From the third session onward, the heterogeneity scores again went down to the subject's average score. But the most improvement we noticed was the total time they took to scan the CXRs was going down while improving their overall accuracy. Compared to the baseline, the total time it took to scan the CXRs for this subject was less than 25% and almost close to an expert.

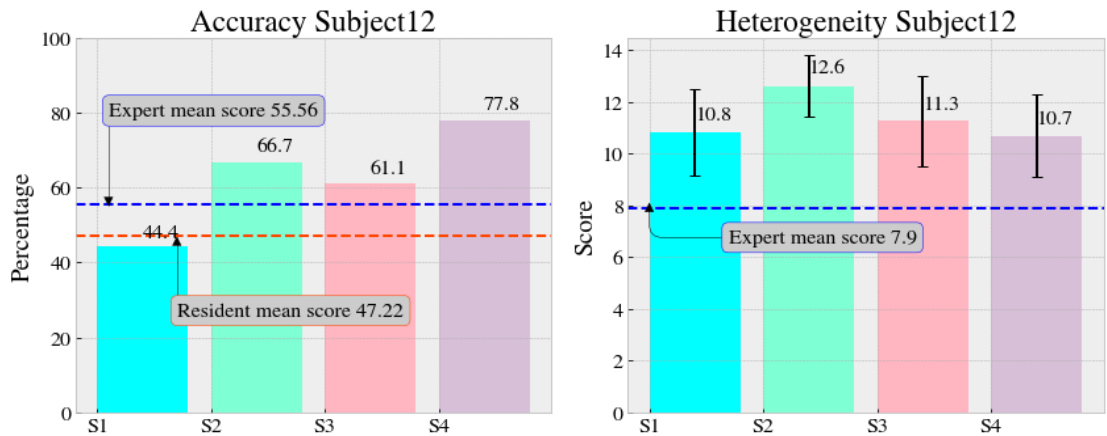

**Figure B.1** - Accuracy and heterogeneity progression for Subject 12: This chart presents Subject 12's accuracy across sessions and heterogeneity progression.

Figures B.1 & B.2 show the detailed statistics of accuracy, heterogeneity, coverage, and total time to scan the CXRs across all the sessions. This supports our hypothesis that supplying proper feedback improves a radiologist's performance. Subject 12 serves as a proof of concept for this educational framework. This resident shows improvement in each of the proposed metrics but interruptions, that never occurred.

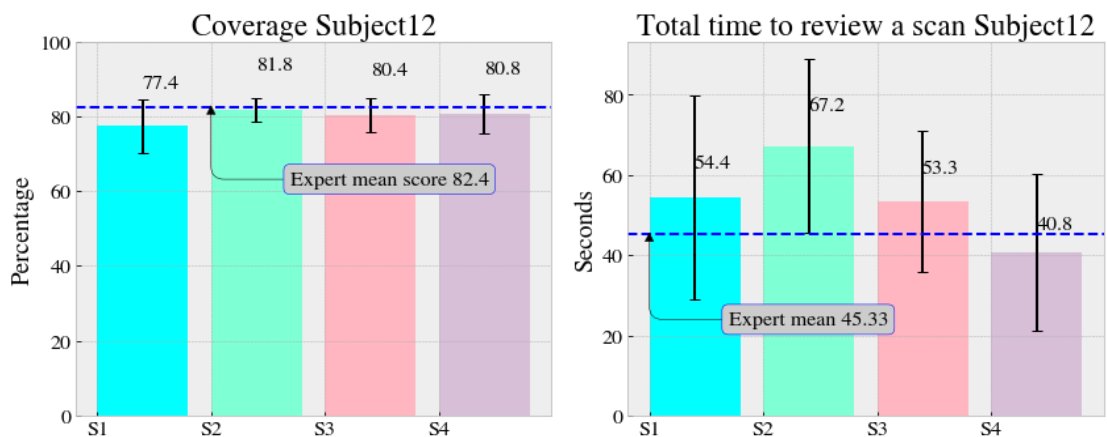

**Figure B.2** - Coverage and total time to review a scan progression for Subject 12: This chart presents Subject 12's Coverage across sessions and the Total time to review a scan progression.

**Appendix C:**

Table C.1 The Detailed mixed-ANOVA Table

| Variable | Source | DF1 | DF2 | F | p-value | η-squared |
|---|---|---|---|---|---|---|
| Accuracy | Group | 1 | 8 | 12.995 | 0.006 | 0.618 |
|  | Session | 3 | 24 | 10.300 | 0.0001 | 0.562 |
|  | Interaction | 3 | 24 | 6.734 | 0.001 | 0.457 |
| Coverage | Group | 1 | 8 | 0.995 | 0.347 | 0.110 |
|  | Session | 3 | 24 | 1.032 | 0.395 | 0.114 |
|  | Interaction | 3 | 24 | 1.649 | 0.204 | 0.170 |
| Heterogeneity | Group | 1 | 8 | 0.831 | 0.388 | 0.094 |
|  | Session | 3 | 24 | 0.571 | 0.639 | 0.066 |
|  | Interaction | 3 | 24 | 0.421 | 0.738 | 0.050 |
| Interruptions | Group | 1 | 8 | 0.839 | 0.386 | 0.094 |
|  | Session | 3 | 24 | 0.124 | 0.944 | 0.015 |
|  | Interaction | 3 | 24 | 1.254 | 0.312 | 0.135 |
| Time | Group | 1 | 8 | 0.616 | 0.454 | 0.071 |
|  | Session | 3 | 24 | 0.766 | 0.524 | 0.087 |
|  | Interaction | 3 | 24 | 0.837 | 0.486 | 0.094 |